\documentclass[
 twocolumn,
 preprintnumbers,
 amsmath,amssymb,
 aps,
 superscriptaddress,
]{revtex4}

\usepackage{times}
\usepackage{stmaryrd}
\usepackage[normalem]{ulem}
\usepackage{comment}
\usepackage[usenames,dvipsnames]{color}
\definecolor{dgreen}{rgb}{0.0, 0.5, 0.0}

\usepackage{color}
\usepackage[dvipdfmx]{graphicx}
\usepackage{bm}
\usepackage{here}
\usepackage{url}
\usepackage{subfigure}
\usepackage{mhchem}
\usepackage{comment}
\usepackage{siunitx}

\newcommand{\bk}[1]{\langle #1\rangle}

\begin{document}
\title{Higher-order Efficiency Bound and Its Application to Nonlinear Nano-thermoelectrics}
%\thanks{A footnote to the article title}%
\author{Takuya Kamijima}
\affiliation{
 Department of Applied Physics, The University of Tokyo, 7-3-1 Hongo, Bunkyo-ku, Tokyo 113-8656, Japan\\}
\author{Shun Otsubo}
\affiliation{
 Department of Applied Physics, The University of Tokyo, 7-3-1 Hongo, Bunkyo-ku, Tokyo 113-8656, Japan\\}
\author{Yuto Ashida}
%\affiliation{
 %Department of Applied Physics, The University of Tokyo, 7-3-1 Hongo, Bunkyo-ku, Tokyo 113-8656, Japan\\}
\affiliation{
 Department of Physics, University of Tokyo, 7-3-1 Hongo, Bunkyo-ku, Tokyo 113-0033, Japan\\}
 \affiliation{
 Institute for Physics of Intelligence, University of Tokyo, 7-3-1 Hongo, Tokyo 113-0033, Japan}
\author{Takahiro Sagawa}
\affiliation{
 Department of Applied Physics, The University of Tokyo, 7-3-1 Hongo, Bunkyo-ku, Tokyo 113-8656, Japan\\}
\affiliation{
 Quantum-Phase Electronics Center (QPEC), The University of Tokyo, 7-3-1 Hongo, Bunkyo-ku, Tokyo 113-8656, Japan\\}

\date{\today}

\begin{abstract}
Power and efficiency of heat engines are two conflicting objectives, and a tight efficiency bound is expected to give insights on the fundamental properties of the power-efficiency tradeoff. 
Here we derive an upper bound on the efficiency of steady-state heat engines, which incorporates higher-order fluctuations of the power. 
In a prototypical model of nonlinear nanostructured thermoelectrics, we show that the obtained bound is tighter than a well-established efficiency bound based on the thermodynamic uncertainty relation, demonstrating that the higher-order terms have rich information about the thermodynamic efficiency in the nonlinear regime.
In particular, we find that the higher-order bound is exactly achieved if the tight coupling condition is satisfied.  
The obtained  bound gives a consistent prediction with the observation that nonlinearity enhances the power-efficiency tradeoff, and would also be useful for various nanoscale engines. 
\end{abstract}

%\pacs{Valid pACS appear here}
%\keywords{Suggested keywords}%Use showkeys class option if keyword

\maketitle
\textit{Introduction.---}
Identifying an ultimate limit on power and efficiency of heat engines is of both fundamental and practical importance.
The tradeoff relation between the power and efficiency, especially in the nonlinear regime, is one of the key issues in nonequilibrium thermodynamics. Historically, Carnot showed that there exists a universal upper bound on the efficiency, $\eta_{\rm C}:=1-\beta_{\rm h}/\beta_{\rm c}$, called the Carnot efficiency, where $\beta_{\rm h}$ and $\beta_{\rm c}$ $(\beta_{\rm h}<\beta_{\rm c})$ are the inverse temperatures of the two reservoirs \cite{callen1998}. For cyclic or steady-state engines, this upper bound is achieved only in the quasi-static and dissipationless limit which requires an infinitely long operation time, leading to a vanishing power. This has been proved in various kinds of models as no-go theorems~\cite{Allahverdyan-Hovhannisyan-Karen-Gevorkian2013,Whitney2014,Brander-Seifert2015,Bauer-Brandner-Seifert2016,Shiraishi-Saito-Tasaki2016}. 
As a pioneering approach to finite time thermodynamics beyond the quasi-static limit, Curzon and Ahlborn showed that the efficiency at maximum power (EMP) is given by $\eta_{\rm CA}:=1-\sqrt{\beta_{\rm h}/\beta_{\rm c}}$ in the endoreversible approximation~\cite{Curzon-Ahlborn1975}.
Such a form of EMP has been shown universal in the linear response regime~\cite{Broeck2005}, where the power and efficiency of thermoelectrics can be characterized only by the figure of merit $ZT$ and the power factor $Q$ \cite{Groot-Sybren-Mazur1984,Benenti-Casati-Saito-Whitney2017, Ashida-Sagawa2021}. 
However, the fundamental tradeoff relation between the power and efficiency has been less established in the nonlinear regime.

It has been recently shown that the efficiency bound (EB) in the nonlinear regime can be obtained from the thermodynamic uncertainty relation (TUR) \cite{Barato-Seifert2015,Gingrich-Horowitz2016,Horowitz-Gingrich-Todd2017,Pietzonka-Ritort-Seifert2017,Pietzonka-Barato-Seifert2016-universal-bound}:
\begin{eqnarray}
\label{TUR}
{\dot{\Sigma}}&\geq& \frac{2\dot{J_d}^2}{\dot{J_d}^{(2)}}=:\widehat{\dot{\Sigma}}_{d,\text{TUR}},
\end{eqnarray}
which gives a lower bound on the (mean) entropy production rate $\dot{\Sigma}$ on the basis of the mean $\dot{J_d}$ and variance $\dot{J_d}^{(2)}$ of the current. 
Here, the variance is rescaled with the observation time so that it gives a non-vanishing value in the long-time limit (the precise definition is given later).
The corresponding EB for steady-state heat engines is given by \cite{Pietzonka-Seifert2018}
\begin{eqnarray}
\label{TUR-based EB}
\eta\leq\frac{\eta_{\rm C}}{1+2P/\beta_{\rm c} P^{(2)}},
\end{eqnarray}
where $P$ and $P^{(2)}$ are the mean and variance of the power, respectively. The EB for molecular motors has also been derived in the same manner \cite{pietzonka-Barato-Seifert2016}. This TUR-based EB has revealed that the Carnot efficiency is not forbidden for a finite power $(P>0)$ provided that the variance of the power diverges $(P^{(2)}\rightarrow\infty)$, which can be regarded as a loophole of the aforementioned no-go theorems and has been demonstrated in  Refs.~\cite{Campisi-Fazio2016,Polettini-Esposito2017,Lee-Park2017,Pietzonka-Seifert2018, Ashida-Sagawa2021} (see also Ref.~\cite{Holubec-Ryabov2018}). 
%\blue{It has been recently pointed out that the diverging variance can be overcome for cyclic engines \cite{Holubec-Ryabov2018}}.
Although the equality of the TUR-based EB can be achieved in the linear response regime, this in general is not the case when the system is far from equilibrium. Thus, there is plenty of room for improving an EB in the nonlinear regime.

In this Letter, we derive a tight bound on the efficiency for steady-state heat engines, which can be achieved even in the nonlinear regime under certain conditions.
A key characteristic of the obtained EB lies in the fact that fluctuations of the power are included as the cumulant generating function to make the bound tighter in the nonlinear regime.
We therefore call the obtained bound  the {\it higher-order} EB,
which can be regarded as a consequence of  the higher-order TUR~\cite{Dechant-Sasa2020} (see Eq.~\eqref{higher-order TUR} below). %\blue{We focus on the average efficiency and ignore the fluctuation of the efficiency \cite{verley-et-2014,Manikandan-Dabelow-Eichhorn-Krishnamurthy2019}}.

To demonstrate the utility of the higher-order EB, we analytically and numerically calculate it for a quantum dot heat engine, a prototypical model of  nanostructured thermoelectrics.
We show that the higher-order terms indeed improve the EB, as the higher-order EB is tighter than the TUR-based EB in this model. 
Furthermore, the EB can even be exactly achieved under the tight coupling condition where the power contains the full information about the total entropy production, allowing one to determine the efficiency only from the information about the power. 
Since it is experimentally feasible to measure relatively low-order ($n\sim5$) cumulants \cite{Gustavsson-2006, Gustavsson-2009}, we can approximate the higher-order EB by truncating its cumulant series up to the $n$-th order. Our numerical calculations show that this truncation can still yield a tighter EB than the TUR-based EB, which can be tested within the current experimental techniques. 
Moreover, the truncated EB only up to $n=4$ estimates the EMP almost exactly.

Our results indicate that the higher-order cumulants of the power (i.e., the power full statistics) beyond its mean and variance are useful to obtain the information of the efficiency in the nonlinear regime.
Specifically, the higher-order EB would be applicable to estimation of the efficiency by using higher-order information of obtained data.
We also see that the higher-order EB is consistent with the observation that nonlinarity enhances the power-efficiency tradeoff~\cite{Schmiedl-Seifert2008,Esposito-Lindenberg-Broeck2009-QD}.

%%%%%%%%%%%%%%%%%%%%%%%%%%%%%%%%%%
\textit{Higher-order efficiency bound.---}
We consider a two-terminal thermoelectric setup where the system is coupled to two reservoirs. The inverse temperatures and chemical potentials of the two reservoirs are denoted by $\beta_{\rm h(c)}$ ($\beta_{\rm h}<\beta_{\rm c}$) and $\mu_{\rm h(c)}$, respectively. The Boltzmann constant is set to be unity. The chemical potential bias $\Delta\mu=\mu_{\rm c}-\mu_{\rm h}>0$ is applied so that the system operates as a heat engine producing the work output by using the temperature bias. Generally, the system in the steady state satisfies the following relation  that is known as the higher-order TUR~\cite{Dechant-Sasa2020}:
\begin{eqnarray}
\label{higher-order TUR}
{\dot{\Sigma}}\geq\sup_{\chi}
\left[\chi {\dot{J_d}}-\lambda_d(-\chi)\right]
=:{\widehat{\dot{\Sigma}}}_d,
\end{eqnarray}
where $\dot{\Sigma}$ is the mean entropy production rate of the thermoelectrics, and $\dot{J_d}$ is a mean generalized current flowing into the system, which is defined by a coefficient $d_{ij}$ specifying the change of an observable in the state transition from $j$ to $i$. $\lambda_d(\chi)$ is the scaled cumulant generating function \cite{Touchette2009} and its derivatives yield the cumulants of the current, that is, $\dot{J_d}^{(n)}
=\frac{\partial^n \lambda_d(\chi)}{\partial\chi^n}|_{\chi=0}\ (n=1,2,\cdots)$. Inequality~\eqref{higher-order TUR} is not exactly the same one derived in Ref.~\cite{Dechant-Sasa2020}, but it is readily obtained in the long-time limit of their result. The equality of the bound~\eqref{higher-order TUR} is always achievable provided that one can choose $d$ arbitrarily. On the contrary, it is not always possible to achieve the equality for a given $d$. 
We leave the derivation of the higher-order TUR in Sec.~A of Supplemental Material, which is mainly based on the Donsker-Varadhan representation \cite{Donsker-Varadhan1983, belghazi2018} of the entropy production.

\begin{figure}
\begin{center}
\includegraphics[width=0.8\linewidth]{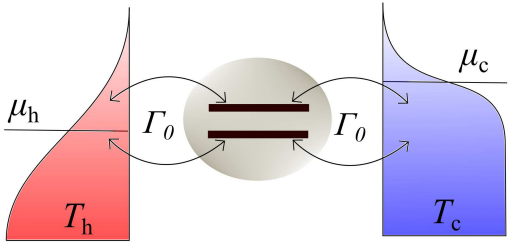}
\caption{Sketch of the two-level quantum dot:
The dot has two energy levels and electrons are flowing from or dissipating into the hot and cold reservoirs. In the sequential tunneling regime and under the wide band approximation, the transition rate is determined by the same coupling strength and the Fermi-Dirac distribution of the level at each reservoir, reflecting upon the occupancy of electrons. In the usual experimental setup, the chemical potentials (equivalently the voltages) are varied while fixing the temperatures. A load to the reservoirs has to be attached to extract the electrical power. For simplicity, we assume the left-right symmetry of the coupling, that is, the coupling strength is identical for both the reservoirs. The capacitance is not installed between the levels, and we assume that there is no interaction between them.}
\label{fig: QD_sketch}
\end{center}
\end{figure}

Using Eq.~\eqref{higher-order TUR}, we derive the EB of the thermoelectrics including the higher-order cumulants of the power $P$. To this end, we adopt the coefficients such that $d_{ij}$ equals $\beta_{\rm c}\Delta\mu(N_i-N_j)$ if the hot reservoir drives the transition $j\rightarrow i$ and $d_{ij}$ is zero otherwise, leading to 
$\dot{J_d}=\beta_{\rm c} P$. $N_i$ is the number of particles lying in the system for the state $i$. Then, the higher-order EB is obtained as
\begin{eqnarray}
\label{higher-order EB}
\eta&=&\frac{\eta_{\rm C}}{\beta_{\rm c}^{-1}P^{-1}\dot{\Sigma} + 1}\nonumber\\
&\leq&\frac{\eta_{\rm C}}{\beta_{\rm c}^{-1}P^{-1}\sup_{\chi}(\chi\beta_{\rm c} P-\lambda(-\chi))+1},
\end{eqnarray}
where $\lambda$ is the scaled cumulant generating function of $\beta_{\rm c} P$.

If we truncate the cumulant series of $\lambda$ at the second term and maximize it in terms of $\chi$, we recover the TUR-based EB with $\chi=2P/\beta_{\rm c}P^{(2)}$. This truncation is justified for the case where the current has the Gaussian distribution or the system is in the linear response regime \cite{Dechant-Sasa2020}. Similarly, we can truncate the series at the $n$-th order term $(n\geq2)$ to approximate the higher-order EB, but it might not be necessarily true that this approximation yields an upper bound of $\eta$ or that the truncated bound gets tighter as including higher-order terms. We discuss this point in Sec.~C of Supplemental Material.
To establish the rigorous relationship between the higher-order EB and the truncated EBs (including the TUR-based EB)  is left as a future issue.

\begin{figure}
\begin{center}
\includegraphics[width=\linewidth]{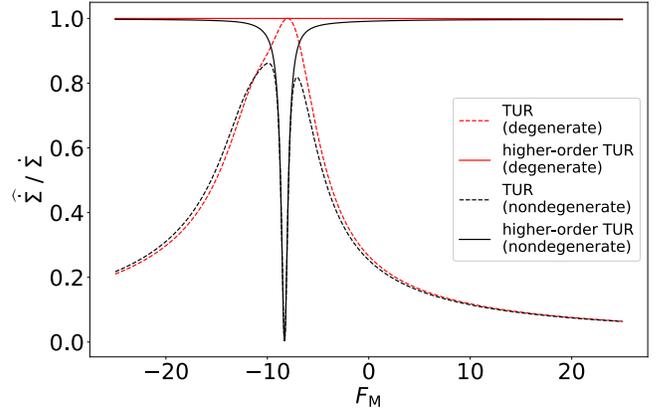}
\caption{Numerical results on the two-level quantum dot:
The estimation of the entropy production rate is plotted according to the TUR (Eq.~\eqref{TUR}) and the higher-order TUR (Eq.~\eqref{higher-order TUR}) with $\dot{J_d}=\beta_{\rm c} P$ for the degenerate case ($x_{\rm h}=2.0$) and the nondegenerate case ($x_{0u}^{\rm h}=x_{d2}^{\rm h}=2.0, x_{0d}^{\rm h}=x_{u2}^{\rm h}=2.2$). The true value, $\dot{\Sigma}$, is obtained by setting $d_{ij}=x_{ij}$. In this example, the calculation is shown outside the heat engine mode $(F_{{\rm M},\text{stop}}\leq F_{\rm M} \leq0)$ for illustration. $F_{\rm H}= 4.0$. $F_{M,\text{stop}}=-8.0$ for the degenerate case.}
\label{fig: TQD_EP}
\end{center}
\end{figure}

\textit{Application to quantum dot heat engines.---}
We demonstrate the higher-order EB (Eq.~\eqref{higher-order EB}) in a quantum dot heat engine. The dot has two energy levels split by the magnetic field (see Fig.\ref{fig: QD_sketch}). We adopt the sequential tunneling approximation so that the transition rate is obtained by the Fermi golden rule \cite{Benenti-Casati-Saito-Whitney2017, Esposito-Lindenberg-Broeck2009-QD}. For the sake of simplicity, we further assume that the coupling strength to the reservoirs is identical $\Gamma_0$ (i.e., wide-band approximation), and there is no interaction between the levels. This strength is small enough to verify the sequential tunneling approximation. The system lies in one of the four distinct states: empty ($0$), one of the levels occupied (${\rm u}$ and ${\rm d}$), and fully occupied ($2$).
The corresponding probability ${\bm p}(t)=[p_0(t)\ p_{\rm u}(t)\ p_{\rm d}(t)\ p_2(t)]^{\rm T}$ obeys the rate equation, $\frac{d}{dt}{\bm p}=W{\bm p}$,
with the transition rate 
$W_{ij}=W_{ij}^{\rm h}+W_{ij}^{\rm c}$. Each nondiagonal element is written as $W_{ij}^{\rm h(c)}:=\Gamma_0f(x_{ij}^{\rm h(c)})\ (i\neq j)$ with $f(x):=1/(1+e^{-x})$, where $x_{ij}^{\rm h(c)}:=\beta_{\rm h(c)}\{(\epsilon_j-\epsilon_i)-\mu_{\rm h(c)}(N_j-N_i)\}$ is the entropy produced in the hot (cold) reservoir as the state changes from $j$ to $i$.  $\epsilon_i$ is the energy of the system in the state $i$. Then, these rates satisfy the local detailed balance condition, $\ln (W_{ij}^{\rm h(c)}/W_{ji}^{\rm h(c)})=x_{ij}^{\rm h(c)}\ (i\neq j)$.
The diagonal elements are determined to satisfy $\sum_i W_{ij}=0$.

The entropy production in the thermoelectric can be written as the bilinear form of the currents \cite{Esposito-Lindenberg-Broeck2009,Groot-Sybren-Mazur1984}:
\begin{eqnarray}
\label{constitutive relation}
\dot{\Sigma}=F_{\rm M} I_{\rm M}^{\rm h}+F_{\rm H} I_{\rm H}^{\rm h},
\end{eqnarray}
where $I_{\rm M}^{\rm h}:=\sum_{i,j}(N_i-N_j)W_{ij}^{\rm h} p_j^{\rm ss}$ and $I_{\rm H}^{\rm h}:=\sum_{i,j}(-x_{ij}^{\rm h})W_{ij}^{\rm h} p_j^{\rm ss}$ with the stationary distribution ${\bm p}^{\rm ss}$ are the particle current and the (normalized) heat current from the hot reservoir, respectively, and $F_{\rm M}:=-\beta_{\rm c}\Delta \mu(<0)$ and $F_{\rm H}:=\beta_{\rm h}^{-1}(\beta_{\rm c}-\beta_{\rm h})(>0)$ are the corresponding conjugate thermodynamic forces. The power and efficiency of the thermoelectric are defined as $P=\Delta\mu I_{\rm M}^{\rm h}$ and $\eta=\beta_{\rm h} P/I_{\rm H}^{\rm h}$, respectively.
Using these definitions, $x_{ij}^{\rm c}$ can be written as $x_{ij}^{\rm c}=x_{ij}^{\rm h}(1+F_{\rm H})+(N_j-N_i)F_{\rm M}$.

When the energy levels are degenerate ($x_{0u}^{\rm h(c)}=x_{0d}^{\rm h(c)}=x_{u2}^{\rm h(c)}=x_{d2}^{\rm h(c)}=: x_{\rm h(c)}$), the tight coupling condition is satisfied, i.e., the particle current and the heat current are proportional to each other ($I_{\rm H}^{\rm h}=x_{\rm h}I_{\rm M}^{\rm h}$) \cite{Benenti-Casati-Saito-Whitney2017}, and Eq.~\eqref{constitutive relation} reduces to $\dot{\Sigma}=FI_{\rm M}^{\rm h}\propto P$ with the effective thermodynamic force $F=F_{\rm M}+x_{\rm h}F_{\rm H}(>0)$. Here, the particle current (or the power) has the full information about the entropy production in a way that all cumulants of it can be obtained from the counterpart of the particle current. The nonlinearity of the system is characterized by $F$ rather than $\eta_{\rm C}$, and $F=0$ corresponds to the stopping force where the power output vanishes. When the levels are nondegenerate and the tight coupling condition is violated, this simple description is no longer valid and the entropy production rate does not vanish in general for the stopping force. Since the number of transported particles cannot determine how much each channel (up or down) is used, the particle current cannot fully capture the entropy production.

According to the counting statistics \cite{Touchette2009}, the scaled cumulant generating function $\lambda(\chi)$ of Eq.~\eqref{higher-order EB} agrees with the principal eigenvalue of the tilted transition matrix, $W_{ij}(\chi)=W_{ij}^{\rm h} e^{\chi\beta_{\rm c}\Delta\mu(N_i-N_j)}+W_{ij}^{\rm c}$. First, in the tight coupling case,
\begin{eqnarray}
&&\lambda(\chi)=2\Gamma_0\Big(-1+\nonumber \\
&&\sqrt{1-f_{\rm h}^+f_{\rm c}^-(1-e^{\chi\beta_{\rm c}\Delta\mu})-f_{\rm h}^-f_{\rm c}^+(1-e^{-\chi\beta_{\rm c}\Delta\mu})}\Big)
\end{eqnarray}
with $f_{\rm h(c)}^\pm=f(\mp x_{\rm h(c)})$.
The mean and variance of the power are obtained from the derivatives of $\lambda$:
\begin{eqnarray}
P&=&\Gamma_0\Delta\mu(f_{\rm h}^+-f_{\rm c}^+), \\
P^{(2)}&=&\Gamma_0(\Delta\mu)^2\Big\{(f_{\rm h}^+f_{\rm c}^- + f_{\rm h}^-f_{\rm c}^+)\nonumber \\ 
&-&\frac{1}{2}(f_{\rm h}^+ - f_{\rm c}^+)^2\Big\}.
\end{eqnarray}
Since $f_{\rm h}^+f_{\rm c}^-=\Lambda(F)e^{F/2}$ and $f_{\rm h}^-f_{\rm c}^+=\Lambda(F)e^{-F/2}$ with $\Lambda(F)=1/(4\cosh(x_{\rm h}/2)\cosh(x_{\rm c}/2))$, the TUR-based EB is written as
\begin{eqnarray}
\label{QD: TUR-based EB}
\eta\leq\frac{\eta_{\rm C}}{\beta_{\rm c}^{-1}P^{-1}\widehat{\dot{\Sigma}}_{\text{TUR}} + 1}
\end{eqnarray}
with
\begin{eqnarray}
\label{QD: TUR}
\widehat{\dot{\Sigma}}_{\text{TUR}}:=\frac{2P^2}{P^{(2)}}
=\Gamma_0\frac{4\Lambda\sinh(F/2)\tanh(F/2)}{1-\Lambda\sinh(F/2)\tanh(F/2)}.
\end{eqnarray}
On the other hand, the maximum of $\chi \beta_{\rm c} P-\lambda(-\chi)$ is obtained by substituting $\chi\beta_{\rm c}\Delta\mu=F$, and the higher-order EB is given by
\begin{eqnarray}
\eta\leq\frac{\eta_{\rm C}}{\beta_{\rm c}^{-1}P^{-1}\widehat{\dot{\Sigma}} + 1}
\end{eqnarray}
with
\begin{eqnarray}
\label{QD: higher-order EB}
\widehat{\dot{\Sigma}}=2\Gamma_0\Lambda F\sinh(F/2)=FI_{\rm M}^{\rm h}=\dot{\Sigma}.
\end{eqnarray}
Thus, the equality of Eq.~\eqref{higher-order EB} is achieved in this case. Up to this point, we have not relied on the heat current to calculate the efficiency.

The standard TUR (Eq.~\eqref{TUR}) and the higher-order TUR (Eq.~\eqref{higher-order TUR}) with $\dot{J_d}=\beta_{\rm c} P$ in the quantum dot heat engine are plotted in Fig.~\ref{fig: TQD_EP}. When the tight coupling condition is satisfied (the degenerate case), the TUR estimation is relatively tight near the stopping force $F_{M,\text{stop}}$ (linear response regime) and the deviation gets larger as the nonlinearity increases, while the higher-order TUR gives the exact estimation for any thermodynamic forces as mentioned above. On the other hand, when this condition is violated (the nondegenerate case), the estimation of the higher-order TUR is no longer exact and its accuracy rapidly drops near the stopping force. This can be understood as follows. The maximum condition, $\chi\beta_{\rm c}\Delta\mu=F$, implies that the counting field $\chi$ plays a role as modifying $d$ to account for the correct entropy production. When the tight coupling condition is violated, in general the entropy production does not vanish for the stopping force (see Eq.~\eqref{constitutive relation}) and $\chi$ cannot recover the entropy production from the vanishing power.

\begin{figure}
\begin{center}
\includegraphics[width=1.0\linewidth]{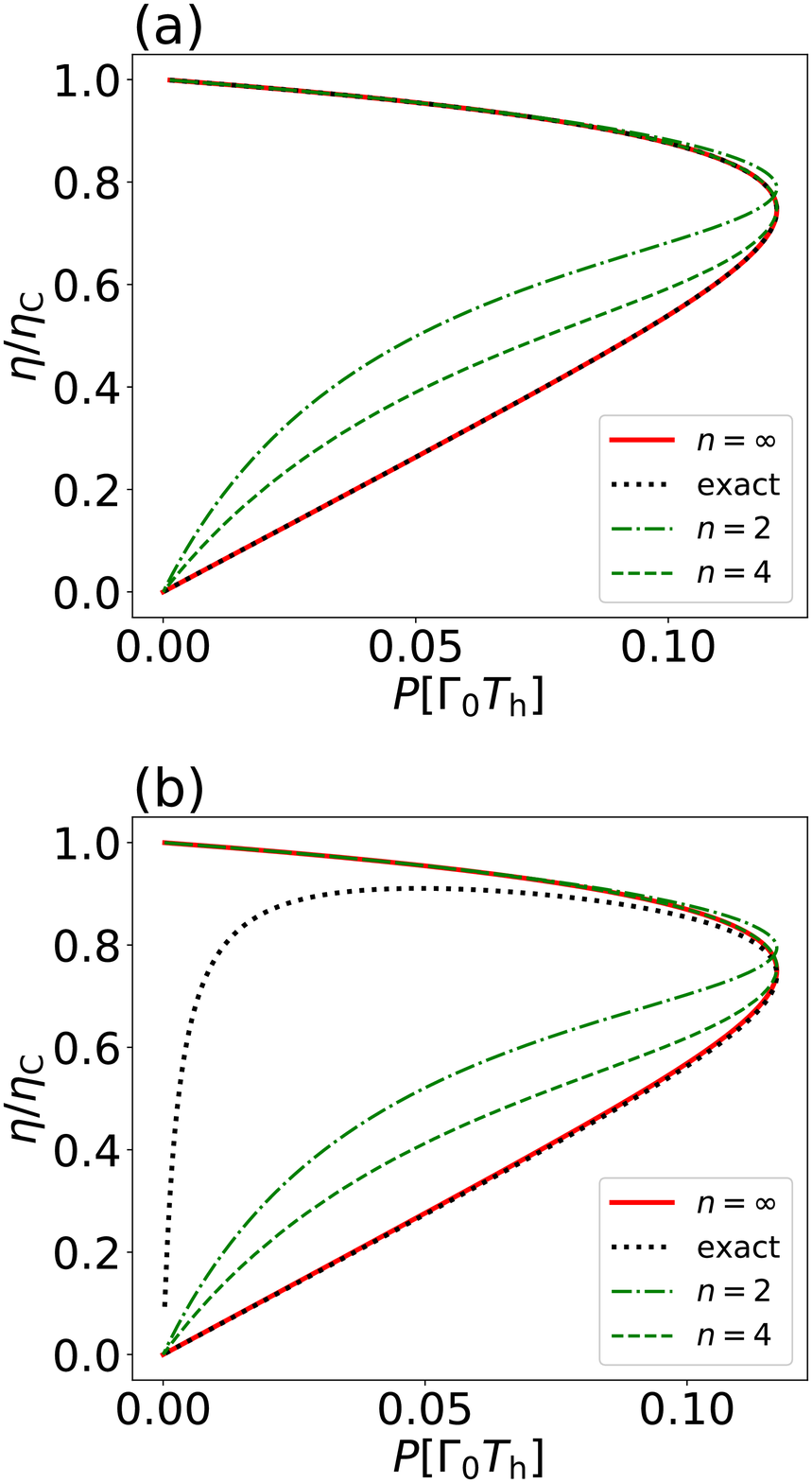}
\caption{Numerical results on the two-level quantum dot: the EBs are plotted according to Eq.~\eqref{higher-order EB}. The estimation of the entropy production is obtained by truncating the cumulant series at the $n$-th order term and maximizing it in terms of $\chi$. The true value, $\eta$, is obtained by dividing the power by the heat current. This heat current is calculated by setting $d$ such that $d_{ij}$ equals $-\beta_{\rm h}^{-1}x_{ij}^{\rm h}$ if the hot reservoir drives  the transition, and equals zero otherwise. (a) The degenerate case, where $F_{\rm H}= 4.0,\ x_{\rm h}=2.0$. (b) The nondegenerate case, where $F_{\rm H}= 4.0,\ x_{0u}^{\rm h}=x_{d2}^{\rm h}=2.0,\ x_{0d}^{\rm h}=x_{u2}^{\rm h}=2.2$.}
\label{fig: TQD_EB}
\end{center}
\end{figure}

The EBs of the quantum dot are plotted in Fig.~\ref{fig: TQD_EB}. These are obtained by truncating the cumulant series of Eq.~\eqref{higher-order EB} at the $n$-th order term and maximizing it in terms of $\chi$. To be accurate, we have to find the local maximum if exists, rather than the global one because of the truncation (see Sec.~C of Supplemental Material). This optimization is always successful for $n=2$ and $n=4$, but not for $n=3$ in this example. $n=2$ and $n=\infty$ correspond to the TUR-based EB, and the higher-order EB, respectively.

For the degenerate case (Fig.~\ref{fig: TQD_EB}(a)), both the power and efficiency increase as $F_{\rm M}(<0)$ decreases from zero. After reaching the maximum power, the efficiency approaches to the Carnot efficiency, and the power vanishes as $F_{\rm M}$ approaches to the stopping force. The truncated EB with $n=4$ is shown tighter than the TUR-based EB and the higher-order EB gives the exact efficiency. 

By contrast, for the nondegenerate case (Fig.~\ref{fig: TQD_EB}(b)), the efficiency also drops as $F_{\rm M}$ decreases to the stopping force after reaching the maximum power. This comes from the nonzero entropy production near the stopping force. Since the higher-order EB cannot take this into account as explained before, there is a large deviation from the exact efficiency near the stopping force. Nonetheless, the higher-order EB estimates the efficiency with high accuracy before reaching the maximum power.

\begin{figure}
\begin{center}
\includegraphics[width=1.0\linewidth]{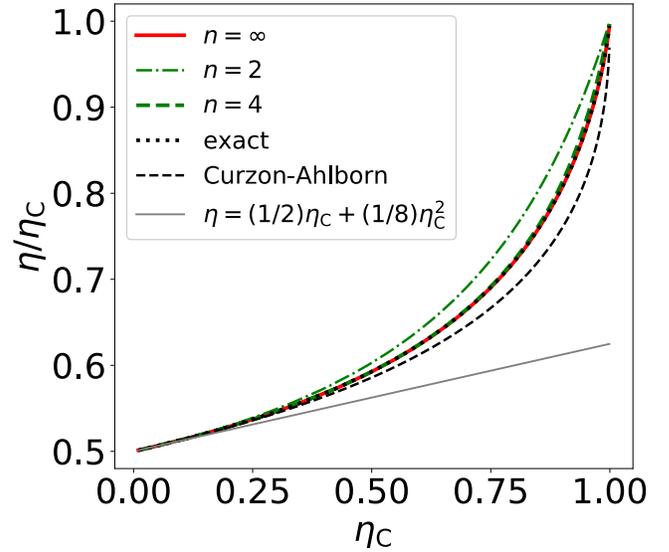}
\caption{Numerical results on the two-level quantum dot:
the efficiency at maximum power and the EBs with the optimal parameters are plotted. These parameters consist of the thermodynamic force $F_{\rm M}$ and the entropy production in the single transition $x_{ij}^{\rm h(c)}$. The levels are assumed to be degenerate in this optimization, because the two channels can be optimized simultaneously under this condition. The EMP $\eta^\ast$ agrees with the one obtained in Ref.~\cite{Esposito-Lindenberg-Broeck2009-QD}, where  nonlinearity enhances the EMP.}
\label{fig: QD_EMP}
\end{center}
\end{figure}

Figure~\ref{fig: QD_EMP} illustrates the efficiency at maximum power (EMP) $\eta^\ast$ and its EB in the quantum dot heat engine. The thermodynamic force $F_{\rm M}$ and the entropy production in the single transition $x_{ij}^{\rm h(c)}$ are optimized to maximize the power output. Since the two channels (up and down) can be optimized simultaneously only for the degenerate case, the tight coupling condition is satisfied in this power optimal condition. Besides, since we assume the left-right symmetry of the coupling, the EMP $\eta^\ast$ asymptotically approaches to 
%the Curzon-Ahlborn efficiency $\eta_{\rm CA}$ in the linear response limit, exhibiting 
the universal form $(1/2)\eta_{\rm C}+(1/8)\eta_{\rm C}^2+\cdots$ proposed in Ref.~\cite{Esposito-Lindenberg-Broeck2009}. This result agrees with the previous research \cite{Esposito-Lindenberg-Broeck2009-QD} where they examined the single-level quantum dot inherently satisfying the tight coupling condition. In the nonlinear regime, 
we see that the EMP is significantly enhanced compared with the above-mentioned universal form and is even larger than the Curzon-Ahlborn efficiency, but exactly agrees with the EB prediction (Eq.~\eqref{higher-order EB}).
We also note that the  truncated EB is much tighter even for $n=4$ than the TUR-based EB ($n=2$).
%, and the EB with the full series ($n=\infty$) agrees with the efficiency.

%%%%%%%%%%%%%%%%%%%%%%%%%%%%%%%%%%
\textit{Discussion.---}
Our higher-order EB can experimentally be tested by using a heterostructured semiconductor nanowire \cite{oDwyer2006, Josefsson-2018, Gustavsson-2006}. The thermodynamic force $F_{\rm M}$ corresponds to the electric voltage applied on the quantum dot, and this can be varied by changing the road resistance $R_{\rm load}$. The power is obtained by measuring the electric current. As for the measurements of the cumulants, the TUR-based EB and the truncated EB with $n=4$ should experimentally be available \cite{Gustavsson-2009, Gustavsson-2006}. For example, we expect that $T_{\rm c} = 1.13 {\rm K}, T_{\rm h} = 1.83 {\rm K}, R_{\rm load}=1.5{\rm M}\Omega$ in Ref.~\cite{Josefsson-2018} are promising parameters to test these EBs at the power optimal condition in the nonlinear regime. We note that, in their experiment, the coupling strength is not small enough to verify the sequential tunneling approximation (and hence they have relied on the RTD theory to calculate the efficiency), and this might cause some deviation from our theory.

Meanwhile, the obtained EB would be applicable to numerical estimation of the efficiency as a higher-order refinement of the TUR-based estimation \cite{Gingrich-Rotskoff-Horowitz2017,Li-et2019quantifying,Pietzonka-Seifert2018,Manikandan-Gupta-Krishnamurthy2020}. 
Specifically, when the cumulants of the steady-state current (e.g., electric current or particle current) are experimentally measured up to the $n$-th order, the estimation of the efficiency can be obtained through Eqs.~\eqref{higher-order TUR} and \eqref{higher-order EB}, by truncating  the scaled cumulant generating function of Eq.~\eqref{higher-order TUR}  and finding a local maximum in terms of $\chi$ (see Sec.~C of Supplemental Material).
%Since the higher-order EB is tighter than the TUR-based EB at least in the thermoelectrics setup, we expect that the higher-order EB in general gives accurate estimation of the efficiency.
We note that our bound gets tightest with the full cumulants without truncation. For estimation of the entropy production, another full cumulant-based estimator has been proposed \cite{Kim-et-2020,Otsubo-Manikandan-Sagawa-Krishnamurthy2020}, while our bound is tighter than it as shown in Sec.~A of Supplemental Material.
Also, the connection to the standard TUR (Eq.~\eqref{TUR}) is clearer in our approach through the truncation.

Finally, we expect that the higher-order EB also gives a tighter bound  for various systems, including Brownian and molecular motors \cite{pietzonka-Barato-Seifert2016, Golubeva-Imparato2012,parrondo2002} and nanosized photoelectric devices~\cite{Rutten-Esposito-Cleuren2009,Pietzonka-Seifert2018}.
This would be plausible because theoretical modelings of these systems  as stochastic processes have common features with thermoelectrics, while explicit demonstrations of the utility of the higher-order EB in various setups are left as future works.

\textit{Acknowledgement.}
T.K. is supported by World-leading Innovative Graduate Study Program for Materials Research, Industry, and Technology (MERIT-WINGS) of the University of Tokyo. Y.A. is supported by JSPS KAKENHI Grant Number JP19K23424. T.S. is supported by JSPS KAKENHI Grant Numbers JP16H02211 and JP19H05796. T.S. is also supported by Institute of AI and Beyond of the University of Tokyo.

%\bibliographystyle{h-physrev}
%\bibliography{higher-order-TUR}
%\bibliography{paper,book} 
%\begin{thebibliography}{60}
%\end{thebibliography}

%\input{main_bib}

%\input{suppl}
\widetext
\pagebreak

\newcommand{\vo}{\upsilon}
\newcommand{\midskip}{\vspace{3pt}}

%\setcounter{figure}{0}
%\def\thefigure{S.\arabic{figure}}
%\setcounter{equation}{0}
%\def\theequation{S.\arabic{equation}}
%%%%%%%%%%%%%%%%%%%%%%%%%%%%%%%%%%%%%%%

\begin{center}
{\large \bf Supplemental Material for  \protect \\ 
  ``Higher-order Efficiency Bound and Its Application to Nonlinear Nano-thermoelectrics'' }\\
\vspace*{0.3cm}
Takuya Kamijima$^{1}$,  Shun Otsubo$^{2}$, Yuto Ashida$^{2,3}$  and Takahiro Sagawa$^{1,4}$ \\
\vspace*{0.1cm}
{$^{1}$Department of Applied Physics, The University of Tokyo, 7-3-1 Hongo, Bunkyo-ku, Tokyo 113-8656, Japan\\}
{$^{2}$Department of Physics, University of Tokyo, 7-3-1 Hongo, Bunkyo-ku, Tokyo 113-0033, Japan\\}
{$^{3}$Institute for Physics of Intelligence, University of Tokyo, 7-3-1 Hongo, Tokyo 113-0033, Japan\\}
{$^{4}$Quantum-Phase Electronics Center (QPEC), The University of Tokyo, 7-3-1 Hongo, Bunkyo-ku, Tokyo 113-8656, Japan\\}
\end{center}

%\begin{thebibliography}{99}
%\end{thebibliography}

\bigskip\noindent
{\bf \large A. Derivation of the higher-order TUR}
\midskip

%\label{section: derivation}
We derive the higher-order TUR (Eq.~\eqref{higher-order TUR} of the main text). We consider a system coupled with multiple reservoirs and assume that it is described as a Markov process and the local detailed balance condition is satisfied. The system evolves along with a time interval $\tau$, and the state of the system is specified at the sequence of time $t_k=k\Delta t$ with $\Delta t=\tau/N\ (k=0,1,\cdots,N)$, giving its trajectory
$\Gamma: i_0\xrightarrow{\nu_0}
i_1\xrightarrow{\nu_1}
\cdots\xrightarrow{\nu_{N-1}}i_N$,
where $i_k$ is the microscopic state at $t=t_k$ and $\nu$ is the index of the reservoir which drives the transition. 
We take the continuous limit $N\rightarrow\infty$ and $\Delta t \to 0$ if necessary.
We assume that the system is in its unique steady state. 
The time-reversal of $\Gamma$ is given by 
$\Gamma^\dagger: i_N\xrightarrow{\nu_{N-1}}
i_{N-1}\xrightarrow{\nu_{N-2}}
\cdots\xrightarrow{\nu_{0}}i_0$.
The state is assumed to be invariant under the time-reversal. The trajectory obeys the probability distribution $P(\Gamma)$, and we define its time-reversal as $Q(\Gamma):=P(\Gamma^\dagger)$. These trajectory distributions take into account the steady state as their initial distributions. An observable $A(\Gamma)$ such as heat or work is measured along trajectories, and its expectation value is denoted as $\langle A\rangle :=\mathbb{E}_P[A]:=\int\mathcal{D}\Gamma A(\Gamma)P(\Gamma)$, where $\mathcal{D}\Gamma$ represents the path integral. Especially, the entropy production can be written as~\cite{Seifert2012}
\begin{eqnarray}
\label{entropy production as KL}
\langle \Sigma\rangle=D_{\rm KL}(P||Q):=\int\mathcal{D}\Gamma P(\Gamma)\ln [P(\Gamma)/Q(\Gamma)],
\end{eqnarray}
where $D_{\rm KL}$ is the Kullback-Leibler divergence of the trajectory distributions.

Then, we introduce the generalized current for the trajectory, $J_d(\Gamma):=\sum_{k=1}^N d_{i_ki_{k-1}}$, where the coefficient $d_{ij}$ determines the change of an observable such as the position and the particle number in the transition from $j$ to $i$. Thus, $J_d(\Gamma)$ counts the net change throughout the trajectory. $d_{ij}$ should be asymmetric, that is, $d_{ij}=-d_{ji}$. The probability distribution of the current is defined as 
\begin{eqnarray}
P_{J_d}(s):=\text{Prob}[J_d(\Gamma)=s]
=\int \mathcal{D}\Gamma\ \delta_{J_d(\Gamma),s}P(\Gamma). 
\end{eqnarray}
The cumulant generating function of the current is given by
\begin{eqnarray}
F_d(\tau,\chi)=\ln Z_d(\tau,\chi)=\ln \sum_{s=-\infty}^{\infty}e^{\chi s}P_{J_d}(s),
\end{eqnarray}
where $\chi$ is called the counting field, and we assume that $F_d(\tau, \chi)$ (and its long-time average $\lambda_d(\chi)$) is differentiable in terms of $\chi$. Then, the $n$-th moment and the $n$-th cumulant of $J_d$ are obtained as $\langle {J_d}^n\rangle=\frac{\partial^n Z_d(\tau,\chi)}{\partial\chi^n}|_{\chi=0}$ and $\langle {J_d}^n\rangle_c
=\frac{\partial^n F_d(\tau,\chi)}{\partial\chi^n}|_{\chi=0}$ respectively.

We are ready to derive the higher-order TUR by using the Donsker-Varadhan representation \cite{Donsker-Varadhan1983, belghazi2018} of Eq.~\eqref{entropy production as KL}:
\begin{eqnarray}
\label{Donsker Varadhan}
D_{\rm KL}(P||Q)\geq\sup_{T\in \mathcal{F}}\left[\mathbb{E}_P[T]-\ln(\mathbb{E}_Q[e^T]) \right],
\end{eqnarray}
where $\mathcal{F}$ is the set of real function, $\{ T(\Gamma) \}$, keeping the expectation values, $\mathbb{E}_P$ and $\mathbb{E}_Q$, finite. The equality is achieved with $T(\Gamma)=\ln (P(\Gamma)/Q(\Gamma))$. 
Substituting $T(\Gamma)=\chi J_d(\Gamma)$ into Eq.~\eqref{Donsker Varadhan}, the following inequality is obtained as
\begin{eqnarray}
\label{unscaled higher-order TUR}
\langle \Sigma\rangle\geq\sup_{\chi}
\left[\chi \langle J_d\rangle-F_d(\tau, -\chi)\right],
\end{eqnarray}
where we used the fact that the Jacobian is invariant under the time-reversal ($\mathcal{D}\Gamma=\mathcal{D}\Gamma^\dagger$) and $J_d(\Gamma^\dagger)=-J_d(\Gamma)$. $\sup_{\chi}$ is added because this derivation is independent of the choice of $\chi$. 
We refer to inequality~(\ref{unscaled higher-order TUR}) as the unscaled higher-order TUR.
This inequality has also been derived in Ref.~\cite{Dechant-Sasa2020} using the Jensen's inequality. Taking the long-time average, we finally obtain Eq.~\eqref{higher-order TUR} with $\dot{J_d}^{(n)}:=\lim_{\tau\rightarrow\infty}\bk{{J_d}^n}_c/\tau
$. For the sake of convenience, we abbreviate the bracket notation to indicate the mean long-time average. In the case of the tight coupling condition, the equality of Eq.~\eqref{higher-order TUR} is achieved, because $\lambda_d$ vanishes at a point where $\chi \dot{J_d}=\dot{\Sigma}$ holds due to the fluctuation theorem \cite{Lebowitz-Joel-Sphon1999, Touchette2009}. %Indeed, $\lambda(F/F_{\rm M})=0$ holds in the quantum dot heat engine. Thus, the equality of Eq.\eqref{higher-order TUR} is trivially achieved suppose we can choose $d$ arbitrarily. By contrast, for a given $d$, the bound is not necessarily achieved like the power in the nondegenerate case.

Note that $\widehat{\dot{\Sigma}}_d$ in Eq.~\eqref{higher-order TUR} is the Legendre transform of $\lambda_d(\chi)$ at $-\dot{J_d}$ for fixed $d$. Since we have assumed the differentiability of $\lambda_d$, $\widehat{\dot{\Sigma}}_d$ becomes a rate function of the (time-averaged) generalized current $J_d/\tau$, that is, $\widehat{\dot{\Sigma}}_d=I_d(-\dot{J_d})=-\chi^*\dot{J_d}-\lambda_d(\chi^*)$ such that $\lambda_d'(\chi^*)=-\dot{J_d}$ with the rate function $I_d(x)$ \cite{Touchette2009}. In light of this, Eq.~\eqref{higher-order TUR} can be also derived from Ref.~\cite{Gingrich-Horowitz2016} with their weakened linear response formula.

If we truncate the cumulant series of Eq.~\eqref{higher-order TUR} and Eq.~\eqref{unscaled higher-order TUR} at the second term and maximize it in terms of $\chi$, we recover Eq.~\eqref{TUR} and the finite-time TUR \cite{Pietzonka-Ritort-Seifert2017,Horowitz-Gingrich-Todd2017}
\begin{eqnarray}
\label{short TUR}
\bk{\Sigma}&\geq& \frac{2\bk{J_d}^2}{\bk{{J_d}^2}_c}.
\end{eqnarray} Again, this truncation can be justified for the Gaussian or the linear response case. If we divide Eq.~\eqref{short TUR} by $\tau$ and optimize $d$ in the short time limit, ${\tau\rightarrow 0}$, the equality is achieved in the Langevin limit and in the equilibrium limit for the Markov jump systems, which is referred to as the short-time TUR \cite{Otsubo-Ito-Dechant-Sagawa2020,Manikandan-Gupta-Krishnamurthy2020,Tan-Tuan-Hasegawa2020, Otsubo-Manikandan-Sagawa-Krishnamurthy2020}.

We note that, on the basis of  another representation of  Eq.~\eqref{entropy production as KL} , called the f-divergence representation~\cite{Nguyen-Wainwright-Jordan2010,Nowozin-Cseke-Tomioka2016,belghazi2018}
\begin{eqnarray}
\label{f-diveegence}
D_{\rm KL}(P||Q)\geq\sup_{T\in \mathcal{F}}\left[\mathbb{E}_P[T]-\mathbb{E}_Q[e^{T-1}] \right],
\end{eqnarray}
the Neural Estimator for Entropy Production (NEEP) \cite{Kim-et-2020}
\begin{eqnarray}
\label{NEEP}
\langle \Sigma\rangle\geq\sup_{d, \chi}
\left[\chi \langle J_d\rangle-Z_d(\tau, -\chi) + 1\right]
\end{eqnarray}
can be obtained by substituting $T(\Gamma)=\chi J_d(\Gamma) + 1$, which optimizes the coefficient $d_{ij}$ as well and considers the higher-order moments of the current. The unscaled higher-order TUR (Eq.~\eqref{unscaled higher-order TUR}) is tighter than
the NEEP for a fixed current, which can be verified by $x-\ln x - 1\geq0\ (x>0)$ with $x=\mathbb{E}_Q[e^{\chi J_d}]$ \cite{belghazi2018}. Therefore, the higher-order EB (Eq.~\eqref{higher-order EB}) is tighter than the EB obtained from the NEEP with $J_d=\beta_c P$.

\bigskip\noindent
{\bf \large B. One-dimensional random walk}
\midskip
%\label{section: RW}

\begin{figure}

\begin{center}
\includegraphics[width=0.9\linewidth]{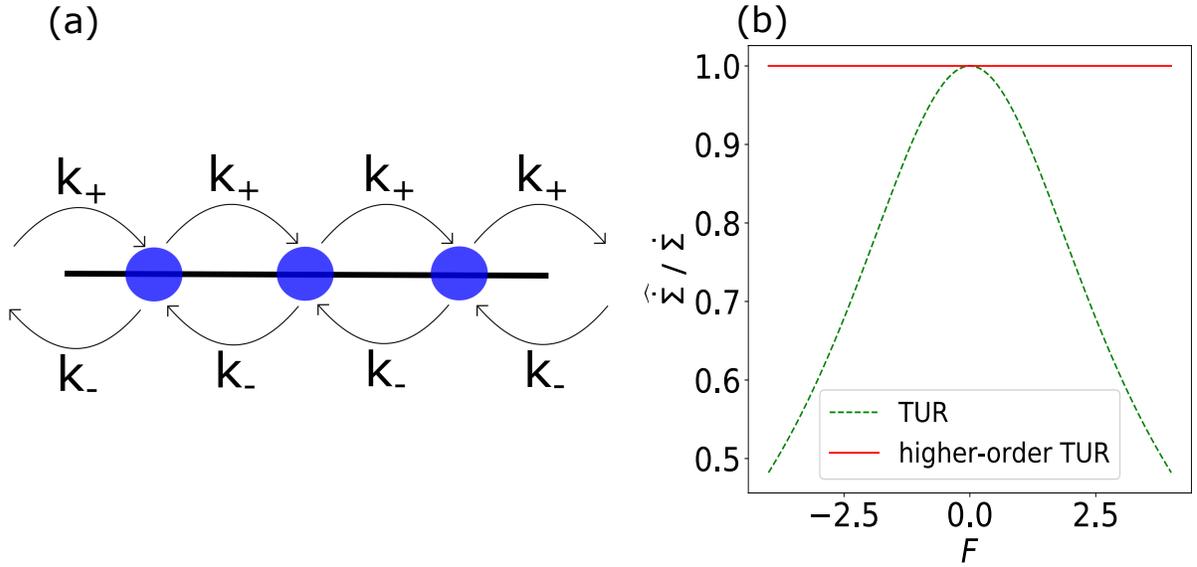}
\caption{(a) The sketch of the one-dimensional random walk.
A single particle is hopping among the infinite sequence of sites. The hopping rate is biased and the particle current is driven by the thermodynamic force $F$. 
(b) Numerical results on the one-dimensional random walk. The estimations of the entropy production based on the TUR (Eq.~\eqref{TUR}) and the higher-order TUR (Eq.~\eqref{higher-order TUR}) are plotted.
}
\label{fig: 1D_RW}
\end{center}
\end{figure}

As another illustration of the higher-order TUR, we consider the one-dimensional random walk in the discrete space and the continuous time, as sketched in Fig.~\ref{fig: 1D_RW}(a). This example has also been investigated  in Ref.~\cite{Dechant-Sasa2020}, while they considered the dynamics in the discrete rather than  continuous time to illustrate the violation of the TUR. In our model, there are infinite sites, and a particle is moving on them forward and back stochastically according to the hopping rate $k_\pm$. We denote the distance between the sites by $d$, which is the coefficient of the current we adopt in this example. $J_d$ is the displacement of the particle and $\dot{J_d}$ is the velocity to the forward direction. Then, the tilted transition matrix is written as
\begin{eqnarray}
W(\chi)=
\begin{bmatrix}
-k_+-k_- & k_+e^{d\chi}+k_-e^{-d\chi} \\
k_+e^{d\chi}+k_-e^{-d\chi} & -k_+-k_-
\end{bmatrix}.
\end{eqnarray}
The principal eigenvalue and the cumulants of the current are straightforwardly calculated as
\begin{eqnarray}
\lambda_d&=&k_+(e^{d\chi}-1) + k_-(e^{-d\chi}-1), \\
\dot{J}_d^{(n)}&:=&\lim_{\tau\rightarrow\infty}\frac{\langle J_d(\tau)^n\rangle_c}{\tau}
=d^{n}(k_+ + (-1)^n k_-),
\end{eqnarray}
respectively. Since the local detailed balance condition is imposed and the hopping is driven by the thermodynamic force $F$, the transition rates must satisfy $k_+/k_-=e^F$. We further assume that the hopping rate is simply written as $k_\pm:= k_0 e^{\pm F/2}$ with the constant rate $k_0$. The entropy production rate is obtained by setting $d=F$:
\begin{eqnarray}
\dot{\Sigma}=\dot{J}_{d=F}=
2k_0 F\sinh(F/2)=k_0\left(F^2+\frac{1}{24}F^4+\cdots\right).
\end{eqnarray}

We are ready to evaluate the TUR (Eq.~\eqref{TUR}) and the higher-order TUR (Eq.~\eqref{higher-order TUR}). The TUR estimator is given by
\begin{eqnarray}
\widehat{\dot{\Sigma}}_{\text{TUR}}
:=\frac{2\dot{J}_d^2}{\dot{J}_d^{(2)}}&=&4k_0\sinh(F/2)\tanh(F/2)\nonumber\\
&=&k_0\left(F^2-\frac{1}{24}F^4+\cdots\right).
\end{eqnarray}
This result is independent of $d$, and the estimation of the entropy production by the TUR is correct only in the lowest order as expected (see Fig.\ref{fig: 1D_RW}(b)). On the other hand, 
\begin{eqnarray}
&&\chi \dot{J}_d-\lambda_d(-\chi)\nonumber\\
&=&2k_0\{d\chi\sinh(F/2)-\cosh(d\chi-F/2)+\cosh(F/2)\} \nonumber\\
&\leq& 2k_0F\sinh(F/2)=\dot{\Sigma},
\end{eqnarray}
where the equality is achieved for $d\chi=F$. Thus, the higher-order TUR estimates the entropy production exactly with this current (see Fig.\ref{fig: 1D_RW}(b)). This maximum condition supports the argument discussed before: the higher-order TUR gives the exact estimation when $\chi$ can modify $d$ to count the entropy production correctly.

%%%%%%%%%%%%%%%%%%%%%%%%%%%%%%%%%%
\bigskip\noindent
{\bf \large C. Truncation of the higher-order EB}
\midskip
%\label{section: truncation}
\begin{figure}
\begin{center}
\includegraphics[width=0.9\linewidth]{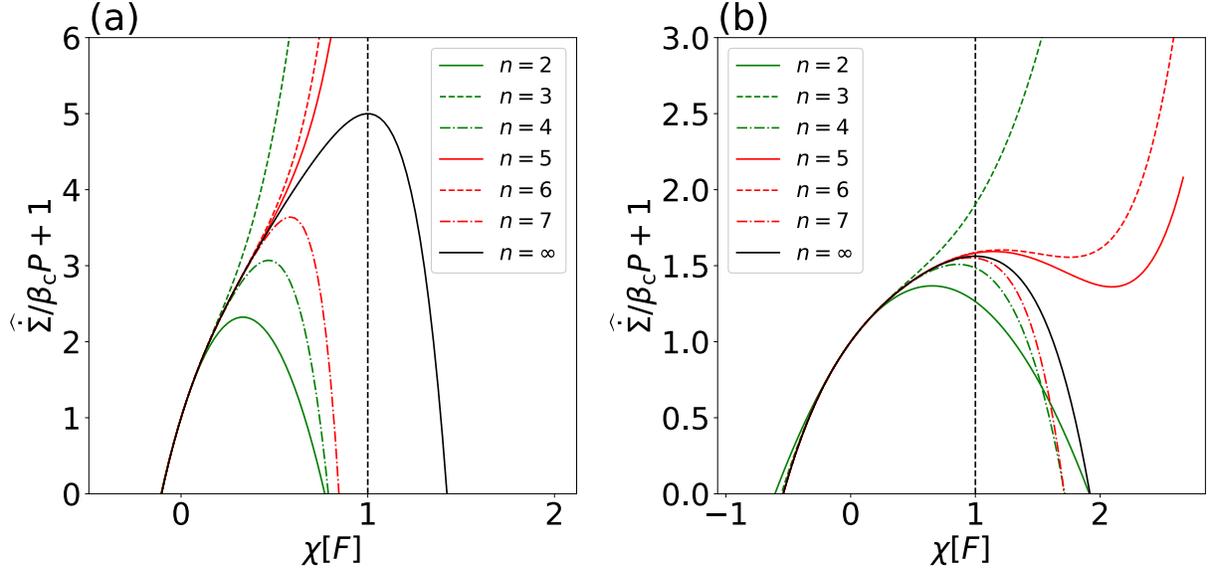}
\caption{Numerical results on the two-level quantum dot in the degenerate case:
the denominators of Eq.\eqref{higher-order EB} before the maximization are plotted. The expansion of $\lambda$ is truncated at the $n$-th order term. The one with the full series ($n=\infty$) has its global maximum at $\chi=F$. $F_{\rm H}=4.0, x_{\rm h}=2.0$. The parameters are given as (a) $F_{\rm M} = -0.20x_{\rm h} F_{\rm H}$, (b) $F_{\rm M} = -0.64x_{\rm h} F_{\rm H}$.}
\label{fig: QD_truncation}
\end{center}
\end{figure}

The maximization procedure of Eq.~\eqref{higher-order EB} is depicted for the degenerate case in Fig.~\ref{fig: QD_truncation}. The coefficient, $d$, is defined such that $\dot{J}=I_{\rm M}^{\rm h}$ for convenience. The denominator, $\beta_{\rm c}^{-1}P^{-1}\widehat{\dot{\Sigma}} + 1$, is plotted with the differently truncated $\lambda$, and the local maxima, if exists, are used to approximate the EB. The EB with the full series ($n=\infty$) has its global maximum at $\chi=F$ according to the analytical calculation.

Let us take a closer look at the plots of $n=5$ and $n=6$ (Fig.~\ref{fig: QD_truncation}(a) and (b)). These illustrate that whether the local maximum exists or not depends on the value of $F_{\rm M}$. Besides, the (local) maximum value is larger than that of $n=\infty$ in Fig.~\ref{fig: QD_truncation}(b), which implies that the truncation does not always yield the upper bound on the efficiency. Furthermore, the value of $n=6$ is slightly larger than that of $n=5$, which shows that increasing $n$ does not always improve the accuracy of this EB.

%%%%%%%%%%%%%%%%%%%%%%%%%%%%%%%%%%
\end{document}